\documentclass[aps,pra,reprint,twocolumn,groupedaddress,nofootinbib]{revtex4-2}
\usepackage{amsmath}
\usepackage{dsfont}
\usepackage{graphicx}
\usepackage{siunitx}
\usepackage{xcolor}

\begin{document}
\title{Adiabatic dynamics in a V-type quantum system by oppositely chirped counterrotating circularly polarized laser pulses}
\author{D. Köhnke, T. Bayer, M. Wollenhaupt}
\email{matthias.wollenhaupt@uol.de}
\affiliation{Carl von Ossietzky university Oldenburg, Institute of Physics, Carl-von-Ossietzky-Straße 9-11, D-26129 Oldenburg}
\date{\today}
\begin{abstract}
Shaped free electron vortices (SEVs) have recently been studied using atomic (1+2) resonance-enhanced multiphoton ionization (REMPI) by oppositely chirped counterrotating circularly polarized (OC-CRCP) femtosecond laser pulses. By transitioning from the perturbative to the non-perturbative REMPI regime, we identify an adiabatic excitation mechanism in a resonant V-type three-level system, termed V-RAP due to its similarities to rapid adiabatic passage (RAP). Experimentally, we observe a pronounced change in the shape of the measured three-dimensional photoelectron momentum distribution (3D PMD), which we trace back to this mechanism via analytical calculations and numerical simulations of the bound state  and ionization dynamics. In V-RAP, the atom adiabatically follows the OC-CRCP field, with the two excited states driven in anti-phase, leading to an adiabatic cancellation of specific ionization pathways and explaining the observed changes in the PMD. In the experiment, we combine supercontinuum polarization pulse shaping to generate OC-CRCP femtosecond laser pulses with velocity map imaging-based photoelectron tomography to reconstruct the 3D PMD. The reconstructed PMDs are decomposed by 3D Fourier analysis into SEVs of different rotational symmetry, revealing a significant enhancement of the $c_6$-symmetric contribution, which is the signature of the V-RAP.
\end{abstract}
\maketitle
\section{Introduction\label{Introduction}}
Tailored ultrashort laser pulses are key to coherently controlling ultrafast quantum dynamics \cite{Brumer:1995:SA:34,Rice:2000:456,Tannor:2007:662}. Among the various pulse shapes routinely generated today by ultrafast pulse shaping techniques \cite{Weiner:2000:RSI:1929,Rabitz:2000:Science:824,Brixner:2001:OL:557,Wohlleben:2005:CPC:850,Silberberg:2009:ARPC:277,Monmayrant:2010:JPB:103001,Wollenhaupt:2012,Misawa:2016:APX:544,Dantus:2017,Qi:2021:APX:1949390}, chirped pulses have played a central role in ultrafast science since the development of the first ultrafast laser sources. Chirped pulses are not only the fundamental basis for the generation of intense femtosecond (fs) laser pulses by chirped-pulse amplification, developed by D. Strickland and G. Mourou, who were awarded the 2018 Nobel Prize in Physics  \cite{Strickland:1985:OC:219,Strickland:2019:RMP:030502}, they have also served as a testbed to demonstrate the adaptive compression of fs-pulses by shaper-based dispersion compensation \cite{Baumert:1997:APB:779,Yelin:1997:OL:1793}. Early on, the chirp was utilized as a control parameter to manipulate ultrafast atomic and molecular dynamics. For example, chirped pulses have been employed for the coherent control of molecular excitation both in the gas phase \cite{Wollenhaupt:2005:ARPC:25} and in the condensed phase \cite{Nuernberger:2007:PCCP:2470}. In the perturbative regime, Cornu spirals emerge in the coherent transients of atomic excitation with resonant chirped pulses \cite{Zamith:2001:PRL:033001,Degert:2002:PRL:203003,Monmayrant:2006:PRL:103002}. In the search for basic mechanisms underlying coherent control, two-dimensional control landscapes of atomic excitation were mapped out by combining the chirp with other physically motivated control parameters \cite{Bayer:2008:JPB:074007,Suchowski:2008:JPB:074008}.
In the non-perturbative regime, intense chirped pulses are the cornerstone of adiabatic passage techniques, enabling robust and efficient population transfer between near-resonantly coupled states \cite{Bergmann:1998:RMP:1003,Vitanov:2001:ARPC:763,Shore:2011:1}. Rapid adiabatic passage (RAP) has been exploited in numerous strong-field control schemes, including the manipulation of spin-orbit wave packet dynamics in atoms \cite{Chatel:2004:PRA:053414}, ladder climbing in atoms \cite{Broers:1992:PRL:2062,Chatel:2003:PRA:041402} and molecules \cite{Maas:1998:CPL:75}, the selective excitation of atomic \cite{Melinger:1992:PRL:2000,Krug:2009:NJP:105051} and molecular states \cite{Melinger:1991:JCP:2210,Bardeen:1995:PRL:3410,Assion:1996:CPL:488} and the control of the Autler-Townes doublet \cite{Autler:1955:PR:703} in the photoelectron spectrum from resonance-enhanced multiphoton ionization (REMPI) \cite{Wollenhaupt:2006:APB:183}. Today the chirp of advanced light sources, such as attosecond laser pulses from high harmonics generation (the so-called ‘attochirp’) \cite{Calegari:2016:JPB:062001,Chang:2016:JOSAB:1081} and ultrashort XUV pulses from free electron lasers \cite{Ilchen:2025:PRR:011001}, is attracting increasing interest in attosecond science. In particular, chirped XUV pulses have recently been \cite{Richter:2024:Nature:337} used to demonstrate strong-field quantum control in the extreme ultraviolet. \\
The use of polarization-tailored ultrashort laser pulses adds another degree of freedom to the coherent control of coherent dynamics in order to exploit the vectorial character of light-matter interactions \cite{Brixner:2004:PRL:208301}. The generation and manipulation of free electron vortices (FEVs) \cite{NgokoDjiokap:2015:PRL:113004,Pengel:2017:PRL:053003} are recent examples of the use of polarization-controlled laser pulses to manipulate light-matter interactions. FEVs are spiral-shaped photoelectron wave packets created by photoionization of atoms \cite{Pengel:2017:PRA:043426,NgokoDjiokap:2016:PRA:013408,Li:2017:COL:120202,Kong:2018:JOSAB:2163,Jia:2019:CPL:119,Zhen:2020:CPL:136885,Armstrong:2019:PRA:063416,Eickhoff:2021:FP:444,Koehnke:2023:NJP:123025} or molecules \cite{Yuan:2017:JPB:124004,Djiokap:2018:PRA:063407,Guo:2021:LP:065301,Bayer:2022:FC:899461,Wang:2023:JOSAB:1749} using sequences of counterrotating circularly polarized (CRCP) pulses. FEVs are characterized by a distinct rotational symmetry in the laser polarization plane, i.e., the number of spiral arms. The creation of FEVs by oppositely chirped CRCP (OC-CRCP) pulses was recently proposed theoretically for the single-photon ionization (SPI) of helium atoms \cite{Strandquist:2022:PRA:043110} and demonstrated experimentally in the (1+2) REMPI of potassium atoms \cite{Koehnke:2024:PRA:053109}. This new class of shaped FEVs (SEVs) exhibits parabolically shaped spiral arms that reverse their sense of rotation at a controlled photoelectron energy and were hence termed `reversible electron spirals' in \cite{Strandquist:2022:PRA:043110}.
In REMPI, the interference of multiple ionization pathways generally leads to the superposition of various SEVs with different rotational symmetries resulting in the total 3D photoelectron momentum distribution (PMD). Specifically, in (1+2) REMPI, a $c_6$-symmetric SEV is superimposed by a $c_4$- and a $c_2$-symmetric SEV. In \cite{Koehnke:2024:PRA:053109}, we reported on the retrieval of the SEVs from the measured PMD using a 3D Fourier analysis method, specifically adapted to the azimuthal symmetry properties of the SEVs. This initial experimental demonstration of the SEV was performed in the perturbative regime. In contrast, in this paper, we study the creation of SEVs by (1+2) REMPI using \textit{intense} OC-CRCP pulses. We observe a distinctive change in the shape of the 3D PMD at the transition from the perturbative to the non-perturbative regime by increasing the laser intensity. Specifically, the SEV with $c_6$ symmetry is significantly enhanced relative to the contributions with $c_4$ and $c_2$ symmetry. This observation is confirmed quantitatively by the 3D Fourier analysis of the measured PMDs.   
Employing an essential three-state model, we analyze the bound state dynamics of the non-perturbative (1+2) REMPI process. Our analysis reveals an adiabatic excitation mechanism in the V-type resonant subsystem, which we term V-RAP. In the paper, we discuss the mechanism in different physical pictures, including the bare state picture and the adiabatic dressed state picture. Analysis of the ionization dynamics shows that the V-RAP mechanism causes destructive interference of certain MPI pathways that contribute to the $c_4$- and the $c_2$-symmetric SEV. The destructive interference explains the suppression of the corresponding symmetry components observed in the 3D PMD. \\
Our results show the potential of SEVs for probing of strong-field induced quantum dynamics. We demonstrate how the 3D Fourier analysis method is used to reveal detailed information about the controlled dynamics enabling us to uncover the V-RAP mechanism.
\section{Physical model}\label{sec:system}
In this section, we describe our physical model for the non-perturbative (1+2) REMPI of potassium atoms by intense OC-CRCP pulses. First, we introduce our notation for the laser electric field of the polarization-shaped pulses. Next, we briefly describe the induced bound state dynamics in the resonant V-type three-level system. Then, we describe the two-photon ionization (2PI) process mapping the bound state dynamics into the photoelectron spectrum.
Finally, we discuss the bound state and ionization dynamics for the transition from the perturbative weak-field regime to the non-perturbative adiabatic interaction regime. 
\subsection{Oppositely chirped CRCP pulses}\label{sec:system:pulses}
The OC-CRCP pulses are generated by spectral phase modulation of a transform-limited input pulse ${\boldsymbol{E^-}(t)=\mathcal{E}_\mathrm{in}(t)e^{-i\omega_0 t}\boldsymbol{e}_y}$, with a real-valued temporal envelope $\mathcal{E}_\mathrm{in}(t)$ of duration $\Delta t$ (full width at half maximum (FWHM) of the intensity) and a carrier frequency $\omega_0$. The following description focuses on the envelope of the laser electric field in the time and frequency domain. The spectrum of the input pulse envelope is denoted by $\tilde{\mathcal{E}}_\mathrm{in}(\omega)=\mathcal{F}[\mathcal{E}_\mathrm{in}(t)](\omega)$.
Next the spectrum of the \mbox{OC-CRCP} pulses is expanded 
in the spherical basis as
\begin{equation}\label{eq:field_w}
	\boldsymbol{\tilde{\mathcal{E}}^-}(\omega)=\tilde{\mathcal{E}}_{+1}^-(\omega)\boldsymbol{e}_{+1}+\tilde{\mathcal{E}}_{-1}^-(\omega)\boldsymbol{e}_{-1}.
\end{equation}
The fields $\tilde{\mathcal{E}}_q^-(\omega)$, with $q=\pm1$, describe the left-handed circularly polarized (LCP) and the right-handed circularly polarized (RCP) pulse component, respectively. Introducing the spectral phase modulation function ${\varphi(\omega)=\frac{1}{2}\,\varphi_2\,\omega^2}$, with the spectral chirp parameter $\varphi_2$, the pulse components are written as
\begin{equation}
	\tilde{\mathcal{E}}_q^-(\omega) = \tilde{\mathcal{E}}_\mathrm{in}(\omega)e^{iq\varphi(\omega)}.
\end{equation}
For a positive chirp parameter, $\varphi_2>0$, the LCP pulse component ($q=+1$) is up-chirped and the RCP pulse component ($q=-1$) is down-chirped. The spherical basis vectors are defined as
\begin{equation}
	\boldsymbol{e}_{\pm1}=\frac{1}{\sqrt{2}}\left(\boldsymbol{e}_y\pm i\boldsymbol{e}_x\right),
\end{equation}
ensuring that for zero chirp, $\varphi_2=0$, the field is linearly polarized parallel to the input pulse (in $y$-direction).
In the time domain, we decompose the envelopes of the chirped pulse components into a envelope $\mathcal{E}_\mathrm{mod}(t)$ and the temporal phase $\zeta(t)$ as
\begin{equation}\label{eq:env_chirp_comp}
	\mathcal{E}^-_q(t)=\mathcal{F}^{-1}\left[\tilde{\mathcal{E}}_q^-(\omega)\right](t)= \mathcal{E}_\mathrm{mod}(t)e^{-iq\zeta(t)}.
\end{equation}
For a real-valued symmetric input pulse spectrum $\tilde{\mathcal{E}}_\mathrm{in}(-\omega)=\tilde{\mathcal{E}}_\mathrm{in}(\omega)$, 
the two pulse components are complex conjugates, i.e., $\mathcal{E}^-_{+1}(t)=[\mathcal{E}^-_{-1}(t)]^\ast$. Otherwise, both pulse components acquire the same additional time-dependent phase due to the asymmetry of the spectrum.
Expanded in the cartesian basis, the temporal field reads
\begin{equation}\label{eq:field_cartesian_t}
	\boldsymbol{\mathcal{E}}(t)=\mathcal{E}_\mathrm{mod}(t)\, \big\{\sin[\zeta(t)]\boldsymbol{e}_x+\cos[\zeta(t)]\boldsymbol{e}_y\big\}.
\end{equation}
The cartesian representation clearly shows, that the pulse is linearly polarized at every time $t$ with a continuously rotating polarization vector. The direction of the polarization vector is determined by the temporal phase $\zeta(t)$, which describes its angle measured relative to the $y$-axis in the polarization plane.
For the specific example of a Gaussian-shaped input pulse, some analytical details are provided in the appendix~\ref{app:gaussian_pulse}. In this case, the temporal phase is a quadratic function of time (see Eq.~\eqref{eq:phase_t}). Therefore, the angular velocity of the rotation varies linearly with time and changes its sign at $t=0$ (see Eq.~\eqref{eq:ang_velo}). 
Figure~\ref{fig1}(a) illustrates the OC-CRCP pulses described by Eqns.~\eqref{eq:field_w} and \eqref{eq:field_cartesian_t} for the example of a Gaussian-shaped input pulse of duration $\Delta t=\SI{6}{fs}$ that is spectrally phase-modulated with a chirp parameter of $\varphi_2=\SI{60}{rad/fs^2}$. The green line shows the real-valued vectorial laser electric field. The orange line showing its envelope highlights the twisting of the linear polarization about the propagation direction.
The rotational characteristics of the OC-CRCP pulses are reminiscent of an optical centrifuge \cite{Karczmarek:1999:PRL:3420,Korobenko:2018:JPBAMOP:203001}. The major difference between both types of pulses lies in their temporal symmetry. While optical centrifuges are asymmetric in time, starting at $t=0$ and accelerating indefinitely under their envelopes, the OC-CRCP pulse is time-symmetric, decelerating during the first part of the pulse ($t<0$) and accelerating during the second part ($t>0$). 
Finally we note that the instantaneous frequency of the OC-CRCP pulse is constant in time and equals the laser central frequency $\omega_0$.
\subsection{Bound-state dynamics}\label{sec:system:bound}
\begin{figure}
	\includegraphics[width=\linewidth]{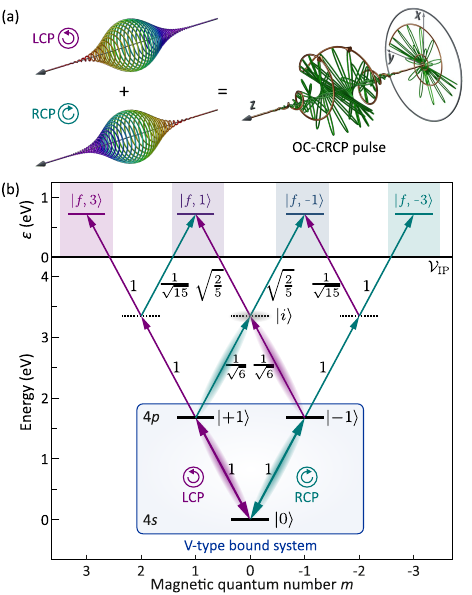}
	\caption{Physical system. (a) Illustration of a Gaussian-shaped OC-CRCP pulse resulting from an input pulse of duration $\Delta t=\SI{6}{\femto\second}$ spectrally phase-modulated with a chirp parameter of $\varphi_2=\SI{60}{\femto\second^2}$. The total polarization-twisted pulse (right-hand side) arises from the superposition of an up-chirped LCP and a down-chirped RCP pulse component (left-hand side).  
		(b) (1+2) REMPI scheme of the potassium atom interacting with an OC-CRCP pulse. 
		The blue box shows the resonantly driven V-type three-state bound system.
		The bound-state dynamics is mapped via different 2PI pathways into four angular momentum partial wave packets, whose pairwise interference gives rise to SEVs of $c_2$, $c_4$ and $c_6$ symmetry.
		\label{fig1}}
\end{figure}
The excitation scheme for OC-CRCP pulses interacting with potassium atoms is shown in Fig.~\ref{fig1}(b). The LCP and RCP pulse components resonantly couple the ground state $|0\rangle=|4s,m=0\rangle$ to the degenerate excited states $|{+1}\rangle=|4p,1\rangle$ and $|{-1}\rangle=|4p,-1\rangle$, respectively. The three states form a V-type linkage pattern \cite{Vitanov:2001:ARPC:763,Shore:2011:1}, as indicated by the blue box in Fig.~\ref{fig1}(b). Choosing $\varphi_2>0$, the state $|{+1}\rangle$ is excited by the up-chirped pulse component $\mathcal{E}^-_{+1}(t)$, while the state $|{-1}\rangle$ is excited by the down-chirped pulse component $\mathcal{E}^-_{-1}(t)$. To describe the light-induced dynamics, we start with the time-dependent Schrödinger equation (TDSE) for the state amplitudes $\boldsymbol{c}(t)=[c_0(t),c_{+1}(t),c_{-1}(t)]^T$ in the interaction picture \cite{Shore:2011:1}. Assuming resonant excitation, $\delta=\omega_0-\omega_{4s\rightarrow4p}=0$, and applying the rotating wave approximation (RWA), the TDSE reads:
\begin{equation}\label{eq:TDSE_int_pic}
	i\hbar\frac{d}{dt}\begin{pmatrix}
		c_0\\
		c_{+1}\\
		c_{-1}
	\end{pmatrix} = -\hbar\begin{pmatrix}
		0	& \Omega(t) & \Omega^\ast(t) \\
		\Omega^\ast(t)	& 0 & 0 \\
		\Omega(t)	& 0 & 0
	\end{pmatrix}\begin{pmatrix}
		c_0\\
		c_{+1}\\
		c_{-1}
	\end{pmatrix},
\end{equation}
with the complex-valued Rabi-frequency ${\Omega(t)=\Omega_0(t)e^{i\zeta(t)}}$. Its real-valued envelope is given by $\Omega_0(t) = \frac{\mu_{\pm1}}{2\hbar} \mathcal{E}(t)$, where we dropped the index 'mod' of the modulated field envelope for brevity. The dipole matrix elements are $\mu_{\pm1}=\langle {\pm1}|\boldsymbol{\mu}|0\rangle$. In writing  the coupling elements of the Hamiltonian as complex conjugates, we additionally assumed the spectrum of the input pulse $\mathcal{E}_\mathrm{in}(\omega)$ to be symmetric, as discussed in Sec.~\ref{sec:system:pulses} (cf. Eq.~\eqref{eq:env_chirp_comp}). Starting from the initial condition $\boldsymbol{c}(-\infty) = (1,0,0)^T$ and applying the short-time evolution operator $\mathcal{U}(t;\delta t)$ specified in Eq.~\eqref{eq:short_time_propagator} of the Appendix, we infer two general characteristics of the dynamics from the Hamiltonian. First, the ground state amplitude $c_0(t)$ is always real-valued. Second, the excited state amplitudes $c_{+1}(t)$ and $c_{-1}(t)$ are negative conjugates of one another, i.e., $c_{+1}(t)=-c_{-1}^\ast(t)$ throughout the interaction. \\
In the weak-field regime, an approximate solution of the TDSE is obtained by first order time-dependent perturbation theory, yielding the excited state amplitudes $c_q(t)$, with $q=\pm1$, as
\begin{equation}\label{eq:solution_weak_field}
	c_q(t)=i\,\frac{\mu_q}{2\hbar}\intop_{-\infty}^t \mathcal{E}(t')e^{-iq\zeta(t')}dt'.
\end{equation}
In the strong field regime, which is characterized by non-perturbative interaction dynamics, analytic solutions are generally not readily available. However, if the time evolution is adiabatic \cite{Vitanov:2001:ARPC:763}, then the total system, consisting of the V-type bound system dressed by the OC-CRCP pulse, remains in an eigenstate throughout the entire interaction. In this case, the TDSE can be solved in the dressed state picture, as we show in the Appendix~\ref{app:bound_dynamics}.
The condition under which the time evolution is adiabatic is also derived in the Appendix~\ref{app:bound_dynamics}. Expressed by the Rabi frequency $\Omega_0(t)$ and the instantaneous detuning $\Delta(t)=\dot{\zeta}(t)$, 
the adiabatic condition for the degenerate V-system reads
\begin{equation}\label{eq:adia_cond_main}
	\big|\dot{\Omega}_0\Delta-\Omega_0\dot{\Delta}\big|\ll \left(2\Omega_0^2+\Delta^2\right)^{3/2}.
\end{equation}
The only difference to the adiabatic condition obtained for a resonantly driven two-state system \cite{Vitanov:2001:ARPC:763} is the additional factor 2 on the right-hand side. By introducing the dimensionless parameter 
\begin{equation}
	\alpha(t)=\frac{\big|\dot{\Omega}_0\Delta-\Omega_0\dot{\Delta}\big|}{ \left(2\Omega_0^2+\Delta^2\right)^{3/2}},
\end{equation}
we obtain a quantitative measure of the transient adiabaticity of the interaction. 
If $\alpha(t) \ll 1$ is satisfied for all times $t$, then the time evolution is fully adiabatic and the system remains in the initially populated dressed state throughout the interaction. The selective population of dressed states, briefly termed SPODS, is a fundamental strong-field control mechanism \cite{Wollenhaupt:2005:JOB:S270,Wollenhaupt:2006:APB:183,Bayer:2016:ACP:235}, which in the context of adiabatic quantum dynamics manifests as  adiabatic following \cite{Vitanov:2001:ARPC:763,Shore:2011:1,Huang:2017:OC:196}. After solving the TDSE in the dressed state picture, we transform the solution back to the interaction picture to obtain the adiabatic bare state solution
\begin{equation}\label{eq:solution_int_pic_main}
	\boldsymbol{c}(t)=\frac{1}{\sqrt{2}}\begin{pmatrix}
		-\sqrt{2}\cos[\Theta(t)] \\
		\sin[\Theta(t)]\,e^{-i\zeta(t)} \\
		-\sin[\Theta(t)]\,e^{i\zeta(t)}
	\end{pmatrix}.
\end{equation}
The mixing angle $\Theta(t)$ is defined by the relation
\begin{equation}\label{eq:mixing_angle}
	\tan[\Theta(t)]=\frac{\sqrt{2}\,\Omega_0(t)}{\Delta(t)}.
\end{equation}
The bare state solution for the adiabatic dynamics in Eq.~\eqref{eq:solution_int_pic_main} showcases the general characteristics mentioned above, i.e., the ground state amplitude $c_0(t)$ is strictly real-valued and the excited state amplitudes $c_{+1}(t)$ and $c_{-1}(t)$ are negative conjugates of one another, for all times $t$. Moreover, the solution reveals that, under adiabatic conditions, the excited state amplitudes $c_{+1}(t)$ and $c_{-1}(t)$ synchronize in-phase and anti-phase with the corresponding pulse components, respectively\footnote{For $\varphi_2<0$, the role of the excited states is exchanged.}. These two simultaneous realizations of adiabatic following, well-known from the traditional RAP scheme \cite{Vala:2001:OE:238,Bayer:2016:ACP:235}, are essential characteristics of the present scenario.
RAP is commonly implemented by driving a two-level system with an intense chirped pulse. Similarly, the adiabatic dynamics in the V-system, consisting of two two-level systems coupled by a common ground state, are driven by the chirped components of the  OC-CRCP pulse.
Owing to the similarities  to RAP, we refer to the adiabatic dynamics in the V-system as V-RAP.
If the input pulse envelope $\mathcal{E}_\mathrm{in}(t)$ is temporally symmetric, i.e. $\mathcal{E}_\mathrm{in}(-t) = \mathcal{E}_\mathrm{in}(t)$, then, unlike in the traditional RAP scheme, the amplitudes also exhibit time reversal symmetry\footnote{Note that for the inversion of Eq.~\eqref{eq:mixing_angle} the two-argument arc-tangent needs to be applied. Hence, one obtains $\Theta(-t)=-\Theta(t)+\pi$ and therefore $\sin[\Theta(-t)]=\sin[\Theta(t)]$.}
$c_q(-t)=c_q(t)$. 
Consequently, the atom is robustly steered back into its ground state by the end of the interaction. This process is known as coherent population return \cite{Vitanov:1995:JPB:1905,Shore:2011:1}.
However, we emphasize that the V-RAP mechanism (adiabatic following accompanied by coherent population return) is also effective for temporally asymmetric input pulse shapes.
\subsection{Ionization dynamics}\label{sec:system:ionization}
In the experiment, the bound state dynamics is mapped into the photoionization continuum by simultaneous 2PI from the excited bound states $|q_0\rangle=|{\pm1}\rangle$. The different 2PI pathways, opened up by the mixing of LCP and RCP photons from the two OC-CRCP pulse components, are indicated in Fig.~\ref{fig1}(b) by one-headed violet and turquoise arrows. Referring to the propensity rules for electric dipole transitions \cite{Fano:1985:PRA:617}, we consider the $\Delta\ell=+1$ transitions. Then, 2PI creates photoelectron partial wave packets in the four different angular momentum continua $|\varepsilon f,\pm1\rangle$ and $|\varepsilon f,\pm3\rangle$. We characterize each ionization pathway by a vector $\boldsymbol{q}=(q_1,q_2)$ where the first (second) component indicates the circularity ($\pm1$) of the first (second) absorbed photon. \\
The two continua $|\varepsilon f,\pm3\rangle$ are each coupled to the bound states by a single 2PI pathway, characterized by the absorption of either two LCP photons from state $|{+1}\rangle$ [$\boldsymbol{q}=(+1,+1)$] or two RCP photons from state $|{-1}\rangle$ [$\boldsymbol{q}=(-1,-1)$]. Therefore, these continua are not subject to multi-pathway interference. In contrast, the two continua $|\varepsilon f,\pm1\rangle$ are each coupled to the bound states by three different 2PI pathways. For example, the continuum $|\varepsilon f,1\rangle$ is coupled to the $|{+1}\rangle$ state by two pathways involving the mixing of one LCP photon and one RCP photon [$\boldsymbol{q}=(+1,-1)$ and $(-1,+1)$], and to the $|{-1}\rangle$ state by an additional pathway involving the absorption of two LCP photons [$\boldsymbol{q}=(+1,+1)$]. \\
If the coupling to the continuum is sufficiently weak, the energy-dependent amplitudes of the photoelectron partial wave packets created along the different 2PI pathways are described by second-order time-dependent perturbation theory \cite{Meshulach:1998:Nature:239,Chatel:2003:PRA:041402,Wollenhaupt:2003:PRA:015401,Eickhoff:2021:PRA:052805}. In general, the photoelectron amplitude created by 2PI from the state $|q_0\rangle$ via the pathway $\boldsymbol{q}=(q_1,q_2)$ is given by 
\begin{equation}\label{eq:photelectron_amp_main}
	a_{q_0}^{(q_1,q_2)}(\varepsilon)=\frac{1}{(i\hbar)^2}\int\limits_{-\infty}^\infty \mathcal{I}_{q_0}^{(q_1,q_2)}(t)e^{i(\varepsilon-\varepsilon_0)\frac{t}{\hbar}}\,dt,
\end{equation}
with the pathway-specific, time-resolved (1+2) REMPI amplitude 
\begin{equation}
	\mathcal{I}_{q_0}^{(q_1,q_2)}(t) = \mu_{q_0}^{(q_1,q_2)} c_{q_0}(t)\mathcal{E}_{q_1}^-(t)\mathcal{E}_{q_2}^-(t).
\end{equation}
Assuming adiabatic interaction conditions, we insert the excited state amplitudes $c_{q_0}(t)$ from Eq.~\eqref{eq:solution_int_pic_main} and make use of the electric fields from Eq.~\eqref{eq:env_chirp_comp} to write the ionization amplitudes explicitly as
\begin{equation}
	\mathcal{I}_{q_0}^{(q_1,q_2)}(t) =  \frac{q_0}{\sqrt{2}}\,\mu_{q_0}^{(q_1,q_2)}\sin[\Theta(t)]\mathcal{E}^2(t)e^{-i(q_0+q_1+q_2)\zeta(t)}.
\end{equation}
Specifically, the ionization amplitude to the continuum $|\varepsilon f,3\rangle$ then reads 
\begin{equation}
	\mathcal{I}_{+1}^{(+1,+1)}(t)=\frac{1}{\sqrt{2}}\sin[\Theta(t)]\mathcal{E}^2(t)e^{- 3i\zeta(t)},
\end{equation}
serving as a reference. Two of the three ionization amplitudes of the continuum $|\varepsilon f,1\rangle$ cancel each other exactly. 
\begin{align}
	\mathcal{I}_{+1}^{(-1,+1)}(t) & = \frac{1}{\sqrt{30}}\sin[\Theta(t)]\mathcal{E}^2(t)e^{-i\zeta(t)} \\
	& \equiv -\mathcal{I}_{-1}^{(+1,+1)}(t).
\end{align}
This implies, that the corresponding 2PI pathways $\boldsymbol{q}=(-1,+1)$ from state $|{+1}\rangle$ and $\boldsymbol{q}=(+1,+1)$ from state $|{-1}\rangle$ interfere destructively throughout the interaction. This phenomenon, which occurs only when the V-RAP mechanism is realized in the bound state system. It will be referred to as adiabatic cancellation (AC) of REMPI pathways in the following. If the AC mechanism applies, then only the pathway $\boldsymbol{q}=(+1,-1)$ from state $|{+1}\rangle$ contributes to the $|\varepsilon f,1\rangle$-type photoelectron partial wave packet, with the ionization amplitude
\begin{equation}
	\mathcal{I}_{+1}^{(+1,-1)}(t) = \frac{1}{\sqrt{30}}\sin[\Theta(t)]\mathcal{E}^2(t)e^{-i\zeta(t)}.
\end{equation}
As a consequence, the photoelectron yield in the continuum $|\varepsilon f,1\rangle$ is significantly suppressed. The analysis of the interplay between bound and ionization dynamics reveals how the V-RAP mechanism is encoded in the $|\varepsilon f,1\rangle$-type partial wave packet. The same conclusions apply to the continua $|\varepsilon f,-1\rangle$ and $|\varepsilon f,-3\rangle$, due to the symmetry of the excitation scheme. \\
The nature of destructive interference, which gives rise to AC, can already be understood at the (1+1) photon level i.e., in the intermediate state $|i\rangle$ at the top end of the diamond-type linkage pattern highlighted in Fig.~\ref{fig1}(b). The state $|i\rangle$ is accessible from the ground state $|0\rangle$ via two different excitation pathways, each of which is composed of one LCP and one RCP photon. Each pathway evolves through one of the excited states $|q_0\rangle$. In the AC case, the excited state amplitudes $c_{q_0}(t)$ effectively mimic the role of the first absorbed photon by adiabatically synchronizing with the corresponding pulse components. However, one state has an in-phase amplitude with the pulse component, while the other has an anti-phase amplitude. Therefore, the absorption of the second photon (of the respective other kind) leads to the same probability amplitudes of both pathways, one of which is phase-shifted by $\pi$, which explains the destructive interference. \\
Our analysis shows that, under adiabatic conditions, the photoelectron partial wave packets created in the continua $|\varepsilon f,\pm1\rangle$ are significantly suppressed relative to those created in the continua $|\varepsilon f,\pm3\rangle$. 
As a result, the amplitudes of the $c_2$- and $c_4$-symmetric SEVs, which consist of superpositions with the $|\varepsilon f,\pm1\rangle$-type partial wave packets, are significantly reduced relative to the amplitude of the $c_6$-symmetric SEV in the PMD. Conversely, the relative enhancement of the $c_6$-symmetric SEV is the hallmark of the V-RAP scenario which should be observable in the experimental PMD.
\begin{figure*}[t]
	\includegraphics[width=\linewidth]{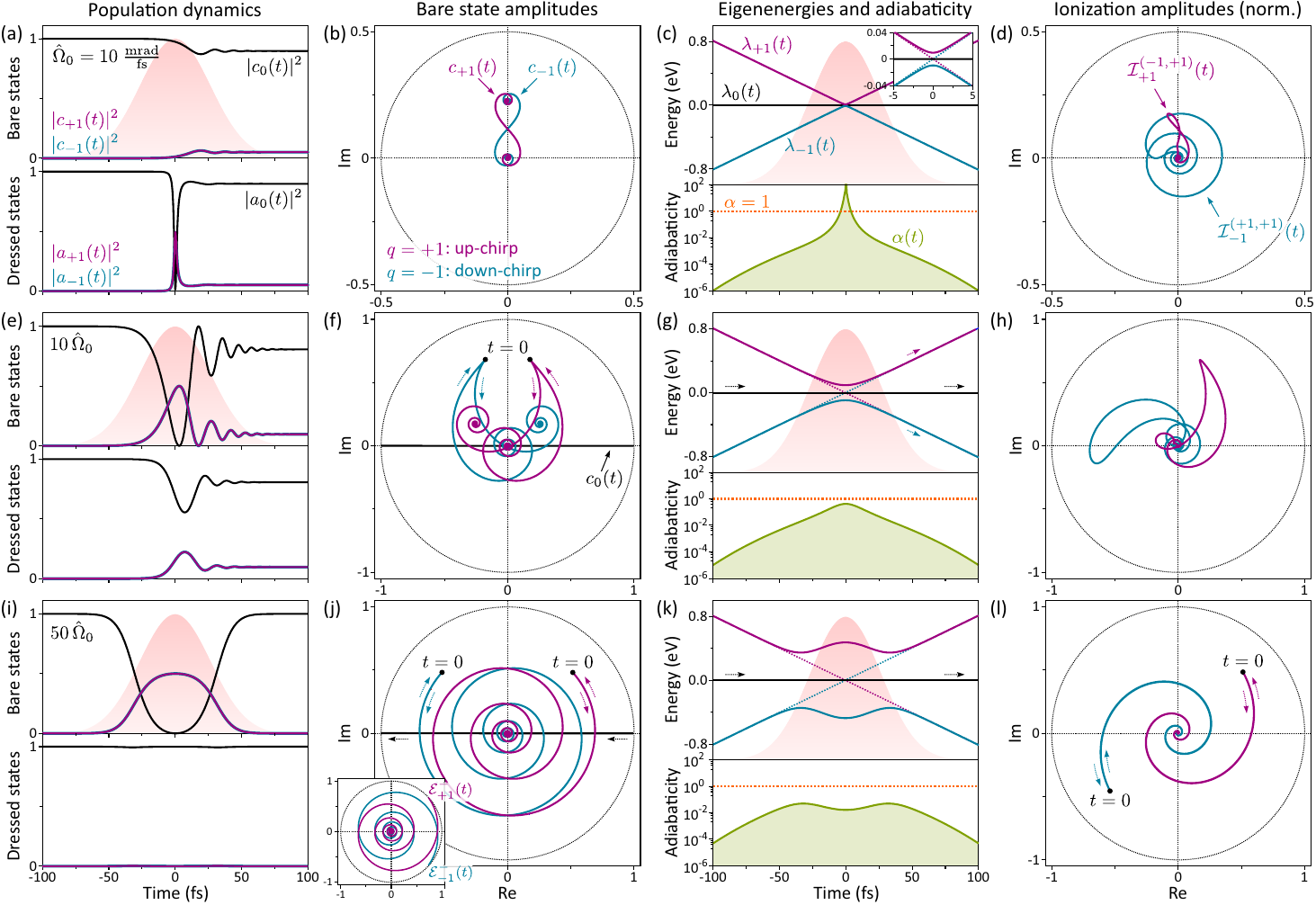}
	\caption{Simulated bound-state and ionization dynamics, induced by a Gaussian-shaped OC-CRCP pulse with an transform-limited duration of $\Delta t=\SI{5}{fs}$ and a spectral chirp parameter of $\varphi_2=\SI{80}{fs^2}$ (up-chirp). Three different interaction regimes, are compared. (a) - (d): The first row shows the perturbative regime with a peak Rabi frequency of $\hat{\Omega}_0=\SI{10}{mrad/fs}$. (e) - (h): The second row shows an intermediate regime with a peak Rabi frequency of $10\,\hat{\Omega}_0$. (i) - (l): The third row shows the adiabatic regime with a peak Rabi frequency of $50\,\hat{\Omega}_0$. In the adiabatic limit, the V-RAP mechanism is observed. V-RAP is characterized by the selective population of the intermediate dressed state throughout the interaction. Because this dressed state is always associated with the atomic ground state, the atom undergoes a coherent population return to its initial state. Transiently, the bare state amplitudes align adiabatically in-phase ($q=+1$) and anti-phase ($q=-1$) with the LCP and RCP pulse component, respectively. As a consequence, the ionization amplitudes $\mathcal{I}_{+1}^{(-1,+1)}(t)$ and $\mathcal{I}_{-1}^{(+1,+1)}(t)$ evolve strictly in anti-phase to one another. Hence they always interfere destructively and cancel each other in the photoelectron wave packet. 
		\label{fig2}}
\end{figure*}
\subsection{From the perturbative to the non-perturbative regime}\label{sec:system:discussion}
In this section, we discuss the bound state and ionization dynamics induced by the OC-CRCP pulses, focusing on the transition from the perturbative to the non-perturbative regime. Figure~\ref{fig2} illustrates three different interaction regimes, organized into different rows. The first row shows the dynamics in the perturbative regime. The peak Rabi frequency is set to $\hat{\Omega}_0=\SI{10}{mrad/fs}$. According to the population dynamics $|c_q(t)|^2$, shown in the top frame of Fig.~\ref{fig2}(a), only 10\% of the ground state population ($q=0$) is transferred to the excited states ($q=\pm1$), which confirms the perturbative interaction conditions. The coherent transients, i.e. the real and imaginary parts of the complex-valued excited-state amplitudes $c_q(t)$, described by Eq.~\eqref{eq:solution_weak_field}, are shown in the parametric plot in Fig.~\ref{fig1}(b). They display the well-known Cornu spirals starting from the origin of the complex plane \cite{Zamith:2001:PRL:033001,Monmayrant:2006:PRL:103002}. 
The top frame of Fig.~\ref{fig2}(c) shows the chirped bare state eigenenergies (thin dotted lines) alongside the corresponding dressed state eigenenergies, labeled as $\lambda_q(t)$ (bold solid lines). As the pulse rises, all energies approach each other. At $t=0$, where the detuning vanishes, the bare state energies intersect whereas the dressed state energies form a bow-tie avoided crossing \cite{Vitanov:2001:ARPC:763,Krug:2009:NJP:105051}. 
Due to the small peak Rabi frequency $\hat{\Omega}_0$, the dressed state energy splitting ${\hbar\Delta\Omega(0)=\sqrt{8}\hbar\hat{\Omega}_0=\SI{20}{meV}}$ is hardly discernible at ${t=0}$. It is only visible in the magnified inset. Accordingly, the adiabaticity parameter $\alpha(t)$, shown in the bottom frame of Fig.~\ref{fig2}(c), peaks sharply at $t=0$ exceeding a value of $100$. At the center of the pulse, the strong non-adiabatic coupling of the dressed states is evident from the rapid change of the dressed state population depicted in the bottom frame of Fig.~\ref{fig2}(a). This observation confirms the non-adiabaticity of the dynamics. As a consequence, the time evolution of the ionization amplitudes $\mathcal{I}_{+1}^{(-1,+1)}(t)$ and $\mathcal{I}_{-1}^{(+1,+1)}(t)$, shown in Fig.~\ref{fig2}(d), is qualitatively different. \\
To illustrate the dynamics in the intermediate field regime (see the second row of Fig.~3), the peak Rabi frequency is increased by an order of magnitude to $10\,\hat{\Omega}_0$. In this case, the interaction is no longer perturbative since the ground state in (e) is transiently depleted (around $t=0$). However, it recovers most of its population after some ringing at the end of the interaction. The excited state amplitudes in (f) form intricate spiral patterns that no longer resemble Cornu spirals, while the ground state amplitude remains strictly real-valued. The dressed state splitting in (g) increases with the field intensity, reaching a value of $\hbar\Delta\Omega=\SI{0.2}{eV}$ at $t=0$. The adiabaticity parameter decreases significantly around $t=0$ and remains smaller than 1 throughout the interaction. However, as illustrated in (e), there is still substantial population transfer between the dressed states, indicating that the condition for fully adiabatic time evolution ($\alpha(t) \ll 1$) is not yet met. On the other hand, the time evolution of the ionization amplitudes, shown in (h), is already quite similar. The two amplitudes form propeller-shaped loops in different directions of the complex plane, visualizing the phase shift between them. \\
The third row of Fig.~\ref{fig2} shows the dynamics in the adiabatic regime. To this end, the peak Rabi frequency is increased to $50\,\hat{\Omega}_0$. As with RAP, the dynamics of the bare state population shown in (i) are characterized by smooth time evolution, with no ringing observed in the ground or excited states. However, unlike in RAP, the dynamics are fully symmetric in time, which is a characteristic of the V-RAP scenario. As a result, the atom returns coherently to the ground state by the end of the interaction. The excited state amplitudes, depicted in the complex plane in (j), start off on spiral-shaped trajectories that are fully analogous to those of the traditional RAP scheme. The amplitudes adiabatically synchronize with the corresponding pulse components in phase ($q=+1$) or anti-phase ($q=-1$), respectively. For direct comparison, the complex envelopes of the pulse components are displayed in the inset. At $t=0$, the trajectories reverse their direction and return to the origin following exactly the same trajectory on which they were excited. This behavior contrasts with that of the traditional RAP scheme, in which the amplitudes do not reverse and continue to increase at the expense of the ground state amplitude until population inversion is reached. Again, the ground state amplitude evolves strictly along the real axis. It crosses the origin at $t=0$ and converges towards $-1$ as $t\rightarrow\infty$. The adiabatic synchronization of the excited state amplitudes with respect to the driving pulse components, including the $\pi$ phase shift, is the central signature of the V-RAP scenario. The dressed state spitting, shown in (k), is now substantial, exceeding $\hbar\Delta\Omega=\SI{1.0}{eV}$ at $t=0$. The adiabaticity parameter is in the order of $10^{-2}$ at all times $t$. This is sufficiently small to ensure fully adiabatic interaction conditions, as confirmed by the dynamics of the dressed state population, shown in (i). There is virtually no population transfer between the dressed states, which implies that the total system remains in its initially populated dressed state throughout the entire interaction. As a result of the adiabatic following in the bound system, the ionization amplitudes in (l) evolve along identical spirals that are rotated by $\pi$ with respect to each other. Therefore, their coherent sum vanishes at all times $t$, giving rise to the AC of the associated 2PI pathways. \\
In Fig.~\ref{figA1} of the Appendix~\ref{app:ionization_dynamics} we show that increasing the field intensity further to $100\,\hat{\Omega}_0$ does not alter the dynamical features of the V-RAP scenario. As is characteristic for adiabatic time evolution, the adiabatic following and the AC mechanism become more robust when higher field amplitudes and/or larger chirp parameters are used \cite{Vitanov:2001:ARPC:763}.
\section{Experimental setup and methods}\label{sec:setup}
\begin{figure*}[t]
	\includegraphics[width=\linewidth]{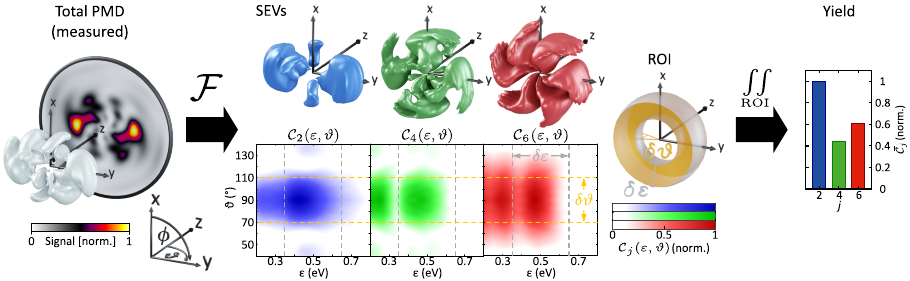}
	\caption{Illustration of the data analysis to extract the yield of the individual SEVs, which are superimposed in the created photoelectron wave packet, from the measured PMD. The PMD shown on the left-hand side is first decomposed into the $c_6$-, $c_4$- and $c_2$-symmetric SEVs using the 3D Fourier analysis method. The SEVs themselves are shown as colored isosurface plots in the top row  of the center frame. The bottom row shows the retrieved 2D amplitudes $\mathcal{C}_j(\varepsilon,\vartheta)$. In the second step, the amplitudes are integrated over the ROI to obtain a scalar measure for the contribution of each SEV to the PMD. The resulting yield is shown as bar plot on the right-hand side.\label{fig3}}
\end{figure*}
The experimental setup and the 3D Fourier analysis method are described in detail in \cite{Koehnke:2024:PRA:053109}. 
Briefly, we combine femtosecond white light polarization pulse shaping \cite{Brixner:2001:OL:557,Weiner:2011:OC:3669,Kerbstadt:2017:JMO:1010} with VMI-based \cite{Eppink:1997:RSI:3477} photoelectron tomography \cite{Wollenhaupt:2009:APB:647,Smeenk:2009:JPB:185402}. The primary light source is a chirped pulse amplification system (\textsc{FEMTOLASERS RAINBOW 500}, \textsc{CEP4 Module}, \textsc{FEMTOPOWER HR CEP}) with a central wavelength of \SI{790}{\nano\meter}, a repetition rate of \SI{3}{kHz} and a pulse energy of \SI{1.0}{\milli\joule}. The infrared laser pulses are used to seed a neon-filled hollow-core fiber at an absolute pressure of \SI{2}{\bar} for the generation of a white light supercontinuum (WLS). Using a home-built $4f$ polarization pulse shaper, based on a dual-layer liquid crystal spatial light modulator (LC-SLM, \textsc{Jenoptik SLM-S640d}), the spectral phases of two orthogonal linearly polarized WLS components are modulated by quadratic spectral phase functions. Using a superachromatic $\lambda/4$ waveplate, the orthogonal linearly polarized components are converted into the CRCP components, resulting in the OC-CRCP output pulse. In addition, a $\lambda/2$ waveplate is used to rotate the output pulse for the tomographic reconstruction of the 3D PMD \cite{Wollenhaupt:2009:APB:647}. The intensity of the OC-CRCP pulses is adjusted using a motorized iris aperture. \\
The polarization-shaped pulses are focused by a spherical mirror ($f=\SI{250}{\milli\meter}$) into the interaction region of a VMI spectrometer filled with potassium vapor from a dispenser source (SAES Getters). The photoelectron wave packets created by atomic MPI are projected onto a micro-channel plate which amplifies the photoelectron signal for visualization on a phosphor screen. The image of the screen is recorded by a charge-coupled device camera (CCD; Lumenera LW165M) using an exposure time of \SI{400}{\milli\second}. To tomographically reconstruct the PMD, the $\lambda/2$ waveplate is rotated in increments of $\SI{2}{\degree}$ from $\SI{0}{\degree}$ to \SI{90}{\degree}, which rotates the PMD effectively by \SI{180}{\degree}. Each projection is measured by averaging over 200 CCD images each exposed for \SI{400}{\milli\second} for every PMD orientation.  From the recorded 45 projections, the 3D PMD is retrieved by application of the Fourier slice algorithm \cite{Kak:1988:1}.\\
Our 3D Fourier analysis method \cite{Koehnke:2024:PRA:053109} is based on the decomposition of the measured 3D PMD $\mathcal{P}(\varepsilon,\vartheta,\phi)$ into azimuthal standing waves. The approach is motivated by the azimuthal parts $e^{im\phi}$ of the created angular momentum partial wave packets, whose pairwise interference give rise to SEVs of different rotational symmetry $j$.  
The PMD is described as a Fourier synthesis  
\begin{equation}
	\mathcal{P}(\varepsilon,\vartheta,\phi) = \sum_{j=0,2,4,6} \mathcal{C}_j(\varepsilon,\vartheta) \cos[j\,\phi + \Phi_j(\varepsilon,\vartheta)].
	\label{eq:Fouriersynthesis}
\end{equation}
The sum runs over the even azimuthal standing waves with a maximum value of $j=6$, because three-photon ionization from the $s$-type potassium ground state creates wave packets with $m=\pm1$ and $m=\pm3$. The azimuthal symmetry of the SEVs resulting from the interference of the partial wave packets is determined by the difference of their magnetic quantum numbers. The decomposition is performed for each energy $\varepsilon$ and polar angle $\vartheta$, yielding the 
two-dimensional (2D) amplitudes $\mathcal{C}_j(\varepsilon,\vartheta)$ and phases $\Phi_j(\varepsilon,\vartheta)$ of the azimuthal standing waves. The procedure is illustrated in Fig.~\ref{fig3} on the example of the PMD created by an intense OC-CRCP pulse with a chirp parameter of $\varphi_2=\SI{-40}{\femto\second^2}$. 
The left-hand side displays the measured total PMD. The individual SEVs of $c_2$, $c_4$ and $c_6$ symmetry are shown in the center frame (top row), together with the retrieved 2D amplitudes $\mathcal{C}_j(\varepsilon,\vartheta)$ (bottom row). \\
According to our discussion in Sec.~\ref{sec:system:ionization}, the essential signature of the V-RAP mechanism encoded in the PMD is the suppression of the $c_2$- and $c_4$-symmetric SEVs relative to the $c_6$-symmetric SEV. The yield of the SEVs is obtained by integrating the retrieved 2D amplitudes over the energy and the polar angle as 
\begin{equation}
	\bar{\mathcal{C}}_j\propto\iint_\mathcal{R}\mathcal{C}_j(\varepsilon,\vartheta)\,\varepsilon^2\sin(\vartheta)\mathrm{d}\varepsilon\mathrm{d}\vartheta.
	\label{eq:SymmetryComponents}
\end{equation}
The region of interest (ROI) denoted by $\mathcal{R}$ is chosen around the maximum photoelectron signal, to minimize the influence of noise from regions with low signal. Along the energy axis, the ROI is centered around the point of reversal of the SEVs \cite{Koehnke:2024:PRA:053109} at $\varepsilon_0=3\hbar\omega_0-\mathcal{V}_\mathrm{IP}\approx\SI{0.5}{\electronvolt}$, with a width of $\delta\varepsilon=\SI{0.3}{\electronvolt}$.
In polar direction, the ROI is centered around $\vartheta_0=\SI{90}{\degree}$, i.e., in the $x$-$y$-plane, with a width of $\delta\vartheta=\SI{40}{\degree}$. The ROI is visualized in the center frame as torus-shaped transparent volume of Fig.~\ref{fig3} and indicated by dashed lines in the 2D plots. The yield of the different SEVs, shown as bar plot on the right-hand side of Fig.~\ref{fig3}, serves as a scalar measure for the quantitative comparison between the PMDs measured in the different interaction regimes. \\
The experiment is supported by numerical simulations to analyze the physical mechanism behind the non-perturbative creation of SEVs by intense OC-CRCP pulses and to validate our analytical model of the V-RAP scenario. The simulations are based on an essential state model including the potassium ground state $4s(m=0)$ and the two excited states $4p(m=\pm1)$. The bound-state dynamics induced by the OC-CRCP pulses are calculated by iteratively solving the TDSE in Eq.~\eqref{eq:TDSE_int_pic} on a discrete temporal grid using the short-time propagation technique \cite{Feit:1982:JCP:412c}. The numerical solution is used to calculate the energy-dependent amplitudes of the created photoelectron partial wave packets in the angular momentum continua $|\varepsilon f,\pm1\rangle$ and $|\varepsilon f,\pm3\rangle$ according to Eq.~\eqref{eq:photelectron_amp_main}. The angular part of the partial wave packets is given by the spherical harmonics $Y_{\ell, m}(\vartheta,\phi)$. The coherent superposition of all partial wave packets describes the total photoelectron wave packet $\psi_e(\varepsilon,\vartheta,\phi)$. The corresponding probability density $|\psi_e(\varepsilon,\vartheta,\phi)|^2$ is compared to the measured PMD $\mathcal{P}(\varepsilon,\vartheta,\phi)$ in energy representation.
\begin{figure*}[htb]
	\includegraphics[width=\linewidth]{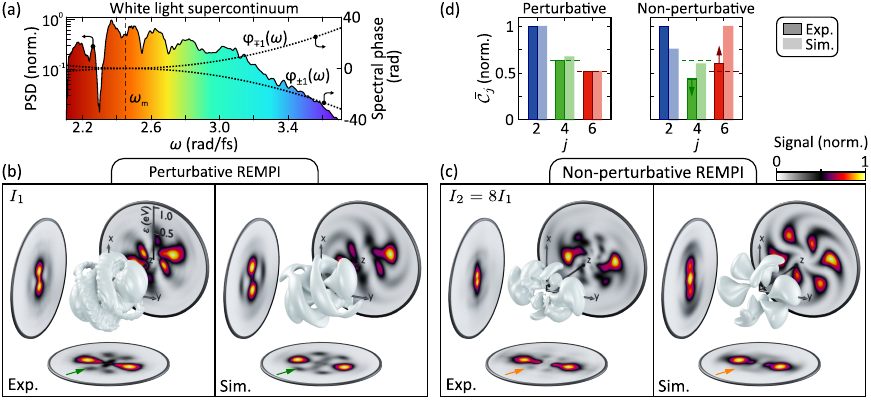}
	\caption{Experimental results. (a) Measured power spectral density (PSD) of the white light supercontinuum used in the experiment, together with the applied quadratic spectral phase functions for a chirp parameter of $|\varphi_2|=\SI{40}{\femto\second^2}$. The bottom row shows tomographically reconstructed PMDs (left frames) and numerically simulated photoelectron densities (right frames) together with their cartesian projections for (b) perturbative and (c) non-perturbative (1+2) REMPI by OC-CRCP pulses. (d) Quantitative evaluation of the experimental and simulation results from (b) and (c) retrieved by 3D Fourier analysis. The bar plots display the yields of the SEVs of different symmetry superimposed in the measured PMD (solid) and simulated density (faded) from perturbative (left) and non-perturbative (right) REMPI.}\label{fig4} 
\end{figure*}
\section{Results}\label{sec:results}
To demonstrate the V-RAP mechanism in the (1+2) REMPI of potassium atoms, we measured PMDs created by OC-CRCP laser pulses in the perturbative and non-perturbative REMPI regime. The PMDs are analyzed and compared both qualitatively and quantitatively regarding their composition of SEVs. \\
The supercontinuum input pulses have a transform-limited pulse duration of about $\Delta t = \SI{6}{\femto\second}$. A measured spectrum of the WLS is shown in Fig.~\ref{fig4}(a) together with the (generic) phase modulation functions $\varphi_{\pm1}(\omega)$ (dotted lines) applied to the orthogonal WLS components.
The spectral chirp parameter was set to $|\varphi_2|=\SI{40}{\femto\second^2}$ and the spectral phase modulation function was applied with respect to the potassium $4s\rightarrow4p$ resonance at $\omega_\mathrm{m}=\omega_{4s\rightarrow4p}=\SI{2.45}{\radian\per\femto\second}$.
The PMDs were recorded at peak intensities of about $I_1\approx\SI{5e11}{\watt\per\centi\meter^2}$, corresponding to perturbative REMPI conditions \cite{Karule:1990:265}, and $I_2\approx8\times I_1$ corresponding to non-perturbative REMPI. Prior to the tomographic reconstruction, the measured 2D projections of the PMDs were symmetrized along the $y$- and the $z$-directions, since PMDs created by single-color MPI are expected to be left-right and forward-backward symmetric with respect to the laser propagation direction \cite{Eickhoff:2021:FP:444}. The redundancy of the experimental data was used to improve the measurement statistics. \\
The experimental results obtained in the perturbative and non-perturbative REMPI regime are presented in Figs.~\ref{fig4}(b) and (c), respectively. The left panels show the measured and reconstructed 3D PMDs in energy representation together with their projections along the cartesian axes. The right panels show the corresponding simulated photoelectron densities. The perturbative PMD was measured using a positive chirp parameter of $\varphi_2=\SI{+40}{\femto\second^2}$. To measure the non-perturbative PMD, we changed the sign of the chirp parameter to $\varphi_2=\SI{-40}{\femto\second^2}$, in order to demonstrate control over the rotational sense of the SEVs \cite{Koehnke:2024:PRA:053109} in addition.\\
We start with a qualitative comparison of the PMDs from perturbative and non-perturbative REMPI.
In the perturbative regime, Fig.~\ref{fig4}(b), the measured PMD exhibits two pronounced main lobes and two ring-type side lobes in between. Hence the PMD strongly resembles that of a rotated $|\varepsilon f,0\rangle$-type wave packet -- as created by three-photon ionization with a linearly polarized pulse \cite{Wollenhaupt:2009:APB:245,Kerbstadt:2017:NJP:103017} -- slightly smeared out in the azimuthal direction. 
In the non-perturbative regime [see Fig.~\ref{fig4}(c)], the polar ring structure of the side lobes disappears, resulting in a more localized 3D density around the polarization plane (the $x$-$y$ plane), reminiscent of the $c_6$ photoelectron vortex reported in \cite{Pengel:2017:PRL:053003}.
This distinctive structural change in the photoelectron distribution is also evident in the simulated photoelectron densities, which accurately reproduce the experimental results. The suppression of the ring lobes can already be clearly seen in the $x$-projection of the PMDs and is, therefore, are directly accessible in the experiment. In the $x$-projection of the perturbative PMD, the rings are clearly visible as isolated signal contributions, marked by the green arrows. In contrast, these signal contributions are barely visible in the $x$-projection of the non-perturbative PMD, as indicated by the orange arrows. The vanishing of the ring lobes and the localization of the non-perturbative PMD around the polarization plane point towards the suppression of $|\varepsilon f,\pm1\rangle$-type contributions, which is the signature of the AC mechanism. The visual inspection of the $z$-projections consistently yields an apparent enhancement of the $c_6$-symmetry of the non-perturbative PMD. The relative suppression of the $c_2$- and $c_4$-symmetric contributions to the PMD indicates the AC in the 2PI process, thus mapping the V-RAP scenario in the bound system. \\
Additionally, the change in sign of the chirp parameter reverses the direction of rotation of the measured PMD and the simulated density. This type of chirp control of the shape of the SEVs has recently been demonstrated in the perturbative regime \cite{Koehnke:2024:PRA:053109}. Our results demonstrate that chirp control of the PMD's rotational sense extends into the non-perturbative regime.\\
Next, we quantify the structural changes in the photoelectron distribution using the 3D Fourier analysis. To this end, we quantitatively investigate the contributions of the individual SEVs to the PMDs in both REMPI regimes by determining the yields $\bar{\mathcal{C}}_j$ as described in Eq.~\eqref{eq:SymmetryComponents}.
The corresponding bar plots in Fig.~\ref{fig4}(d) show the experimental results (solid bars) in direct comparison to the simulated results (faded bars). In the perturbative regime, the results of the experiment and the simulation are in excellent agreement, validating our numerical model. In the non-perturbative regime, the experiment and model are still in reasonable agreement. In both cases, we observe an increase in the yield of the $c_6$-symmetric SEV relative to the yields of the $c_2$- and $c_4$-symmetric SEVs, as indicated by the vertical dashed lines. The numerical model shows a more pronounced relative increase in the $c_6$-symmetric SEV than the experimental results. In general, strong-field effects are more challenging to observe experimentally as they tend to be masked by the averaging of the focal intensity distribution. Besides this general issue, we identify two specific causes of the discrepancies between the experiment and the idealized model. Firstly, the WLS spectrum used in the experiment [see Fig.~\ref{fig4}(a)] is not symmetric. This would have been necessary in order to implement an ideal V-RAP mechanism, as discussed in Sec.~\eqref{sec:system:bound}, relating to the Hamiltonian in Eq.~\eqref{eq:TDSE_int_pic}. Hence, the input pulse already exhibits a time-dependent phase, implying that the two OC-CRCP pulse components coupling the two excited states to the common ground state are not exact complex conjugates. 
While this additional phase does not reduce the effectiveness of the V-RAP mechanism, which is robust with respect to moderate variations of the pulse parameters, it does affect the AC mechanism. The excited state amplitudes synchronize with the different phase dynamics of the pulse components and, hence, are not perfectly in anti-phase. As a consequence, the destructive interference of  2PI pathways discussed above is incomplete.  
Secondly, due to the ultrabroad bandwidth of the WLS, higher-lying bound states of the potassium atom, such as the $3d$- and the $4d$-state, become resonantly accessible \cite{Bayer:2023:PRA:033111}. These states are not considered in our essential state model.  
These intermediate resonances introduce additional ionization phases along the 2PI pathways, which, predominantly affect the AC mechanism rather than the bound state dynamics in the resonant V-system. Despite these limitations, the destructive interference of 2PI pathways leading to the $|f,\pm1\rangle$-continua is clearly observed in the experimental results, and even more so in the numerical simulation. We conclude that the pronounced changes observed in the PMD confirm the V-RAP scenario in the potassium $4s$-$4p$-system.
\section{Conclusion\label{Conclusion}}
In this paper, we studied the creation of SEVs by OC-CRCP femtosecond laser pulses in the (1+2) REMPI of potassium atoms, transitioning from the perturbative to the non-perturbative interaction regime. The OC-CRCP pulses were generated by supercontinuum polarization pulse shaping and the 3D PMDs were measured by VMI-based photoelectron tomography. Applying a 3D Fourier analysis method, we showed that the created  photoelectron wave packet is a superposition of SEVs exhibiting $c_2$, a $c_4$ and a $c_6$ rotational symmetries. 
Comparing the 3D PMDs measured for both the perturbative and non-perturbative REMPI revealed distinctive structural changes in the photoelectron distribution.
Quantitative evaluation of the symmetry contributions yielded a relative increase of the $c_6$-symmetric SEV in the PMD from non-perturbative REMPI. The corresponding suppression of the $c_2$- and $c_4$-symmetric SEVs was traced back to the destructive interference of 2PI pathways in the REMPI by intense OC-CRCP pulses.
An analytical investigation of the non-perturbative (1+2) REMPI process in both the bare state and the dressed state picture revealed that the destructive interference of the ionization pathways is caused by an adiabatic excitation mechanism in the resonant V-type three-level system, termed V-RAP. 
Our analysis of the V-RAP mechanism showed that the excited state amplitudes are adiabatically driven in anti-phase with one another by the OC-CRCP pulse components, resulting in an adiabatic cancellation of specific ionization pathways throughout the interaction. The anti-phase relationship of the excited states accompanied by the coherent population return of the atom into the ground state at the end of the interaction are the central signatures of the V-RAP mechanism. \\
Our results demonstrate the potential of SEVs as powerful and highly sensitive tools for the differential analysis of ultrafast excitation and ionization dynamics in different light-matter interaction regimes.
In the near future, we plan to experimentally and theoretically investigate the non-perturbative bound-state dynamics induced by OC-CRCP pulses. In the experiment, we will use our multichromatic polarization shaping scheme \cite{Koehnke:2023:NJP:123025} to generate an additional ultrashort probe pulse of a different color to perform shaper-based polarization-sensitive pump-probe experiments for the time-resolved background-free imaging of the V-RAP dynamics. In addition, we will carry out numerical quantum dynamics simulations employing our ab initio 2D TDSE model \cite{Bayer:2023:PRA:033111} to shed light on the interaction between the induced vectorial dipole dynamics in the bound system and the shaped vectorial light field at the level of the wave function.
\begin{acknowledgments}
	We gratefully acknowledge financial support from the collaborative research program “Dynamics on the Nanoscale”
	(DyNano) and the Wissenschaftsraum “Elektronen-Licht-Kontrolle” (elLiKo) funded by the Niedersächsische Ministerium für Wissenschaft und Kultur.
\end{acknowledgments}
\appendix
\section{Chirped Gaussian-shaped pulses}\label{app:gaussian_pulse}
The description of the OC-CRCP pulses and the V-RAP scenario in the main text is largely general and valid for a large class of input pulse shapes. For the discussion of specific physical properties and their illustrations, we have used Gaussian-shaped pulses for which an analytical description of second order spectral phase modulation is available. Gaussian-shaped input pulses are characterized by an envelope of the form
\begin{equation}
	\mathcal{E}_\mathrm{in}(t)=\mathcal{E}_0 e^{-\ln{(4)}\left(\frac{t}{\Delta t}\right)^2},
\end{equation}
As detailed for example in \cite{Wollenhaupt:2012}, the modulated field resulting from second order spectral phase modulation is also Gaussian-shaped, with a temporal phase function $\zeta(t)$ given by 
\begin{equation}\label{eq:phase_t}
	\zeta(t)=\chi t^2-\varphi_0.
\end{equation}
The temporal chirp parameter reads
\begin{equation}
	\chi=\frac{1}{2}\,\frac{\varphi_2}{\varphi_2^2+\gamma^2},
\end{equation}
with the abbreviation $\gamma = \frac{\Delta t^2}{\ln(16)}$, and the chirp-dependent constant phase is
\begin{equation}
	\varphi_0=\frac{1}{2}\arctan\left(\frac{\varphi_2}{\gamma}\right).
\end{equation}
The modulated envelope $\mathcal{E}_\mathrm{mod}(t)$ is temporally stretched due to the chirp by a factor of
\begin{equation}
	\frac{\Delta t_\mathrm{mod}}{\Delta t}=\sqrt{1+\left(\frac{\varphi_2}{\gamma}\right)^2}.
\end{equation}
Its peak amplitude is decreased by
\begin{equation}
	\frac{\mathcal{E}_\mathrm{0,mod}}{\mathcal{E}_0}=\frac{1}{\sqrt[4]{1+\left(\frac{\varphi_2}{\gamma}\right)^2}}.
\end{equation}
In the main text, we describe that the OC-CRCP pulses are linearly polarized with a continuously rotating polarization vector (cf. Eq.~\eqref{eq:field_cartesian_t}). The polarization direction relative to the $y$-axis is determined by the temporal phase $\zeta(t)$. For a Gaussian-shaped pulse we obtain the angular velocity of the rotation explicitly as
\begin{equation}\label{eq:ang_velo}
	\dot{\zeta}(t)=\frac{\varphi_2}{\varphi_2^2+\gamma^2}\,t.
\end{equation} 
The angular velocity hence varies linearly in time and changes sign (direction) at $t=0$. The rate of change depends non-linearly on the chirp parameter $\varphi_2$. For a given pulse duration $\Delta t$, the maximum rate of change is obtained for a chirp parameter of $\varphi_2=\gamma$. 
\section{Bound-state dynamics}\label{app:bound_dynamics}
In this appendix, we describe the theoretical basics for our discussion of the V-RAP mechanism in the bound system and its manifestation in the photoelectron spectrum.
In particular, we derive the adiabatic condition that ensures the total interacting system, consisting of the V-type atomic three-level system and the OC-CRCP pulse, remains in an eigenstate throughout the interaction. Different physical pictures are employed to highlight various aspects of the interaction. The perturbative MPI process is considered in the interaction picture. The bound state dynamics are described in a frame rotating with the laser instantaneous frequency, as is common for chirped excitation \cite{Vitanov:2001:ARPC:763}. The adiabatic condition is derived in the corresponding dressed state picture. Under adiabatic conditions, the TDSE decouples in the dressed state picture and is therefore readily solved. Finally, this solution is transformed back into the interaction picture for the calculation of the photoelectron amplitudes. \\
We start with the TDSE for the state amplitudes ${\boldsymbol{c}(t)=[c_0(t),c_{+1}(t),c_{-1}(t)]^T}$ of the V-type three-level system in the interaction picture \cite{Shore:2011:1}. Assuming resonant excitation, $\delta=\omega_0-\omega_{4s\rightarrow4p}=0$, and applying the rotating wave approximation (RWA), the TDSE reads:
\begin{equation}\label{eq:TDSE_int_pic_app}
	i\hbar\frac{d}{dt}\begin{pmatrix}
		c_0\\
		c_{+1}\\
		c_{-1}
	\end{pmatrix} = -\hbar\begin{pmatrix}
		0	& \Omega(t) & \Omega^\ast(t) \\
		\Omega^\ast(t)	& 0 & 0 \\
		\Omega(t)	& 0 & 0
	\end{pmatrix}\begin{pmatrix}
		c_0\\
		c_{+1}\\
		c_{-1}
	\end{pmatrix},
\end{equation}
with the complex-valued Rabi-frequency ${\Omega(t)=\Omega_0(t)e^{i\zeta(t)}}$ and the real-valued envelope $\Omega_0(t)=\frac{\mu_{\pm1}}{2\hbar}\mathcal{E}(t)$. The interaction picture Hamiltonian on the right-hand side of Eq.~\eqref{eq:TDSE_int_pic_app} is denoted by $\mathcal{H}(t)$. \\
By defining the differential pulse area ${\theta(t;\delta t) = \sqrt{2}\Omega_0(t)\delta t}$ and using the abbreviation $\beta(t)=\frac{i}{\sqrt{2}}e^{-i\zeta(t)}$, the short-time evolution operator $\mathcal{U}(t;\delta t) = e^{-i \frac{\delta t}{\hbar} \mathcal{H}(t)}$ is written as (arguments suppressed for better readability)
\begin{equation}\label{eq:short_time_propagator}
	\mathcal{U} = 
	\begin{pmatrix}
		\cos\left(\theta\right)	& -\beta^\ast\sin\left(\theta\right) &  \beta\sin\left(\theta\right) \\
		\beta\sin\left(\theta\right)	&  \cos^2\left(\frac{\theta}{2}\right) & 2\beta^2\sin^2\left(\frac{\theta}{2}\right) \\
		-\beta^\ast\sin\left(\theta\right)	& 2\left(\beta^\ast\right)^2\sin^2\left(\frac{\theta}{2}\right) & \cos^2\left(\frac{\theta}{2}\right)
	\end{pmatrix}.
\end{equation}
One verifies generally that for any state vector $\boldsymbol{c}(t)$ fulfilling $c_0(t)\in\mathds{R}$ and $c_{+1}(t)=-c_{-1}^\ast(t)$, these properties are preserved under the propagation by $\mathcal{U}(t;\delta t)$. Assuming the initial condition $\boldsymbol{c}_0=\boldsymbol{c}(-\infty)=(1,0,0)^T$, the solution to Eq.~\eqref{eq:TDSE_int_pic_app} exhibits these properties in any interaction regime, i.e. irrespective of the pulse parameters. \\
Next we determine the Hamiltonian in the frame rotating with the laser instantaneous frequency by applying the unitary transformation
\begin{equation}
	\mathcal{T}(t)=\begin{pmatrix}
	1	& 0 & 0 \\
	0	& e^{i\zeta(t)} & 0 \\
	0	& 0 & e^{-i\zeta(t)}
	\end{pmatrix}
\end{equation}
which yields:
\begin{align}
	\mathcal{H}' & =\mathcal{T}\mathcal{H}\mathcal{T}^\dagger-i\hbar\mathcal{T}\dot{\mathcal{T}}^\dagger \label{eq:hamiltonian_transf}\\
	& =-\hbar\begin{pmatrix}
		0	& \Omega_0 & \Omega_0 \\
		\Omega_0	& \Delta & 0 \\
		\Omega_0	& 0 & -\Delta
	\end{pmatrix}.
\end{align}
Here we have introduced the instantaneous detuning $\Delta(t)=\dot{\zeta}(t)$. The second term in Eq.~\eqref{eq:hamiltonian_transf} arises due to the transformation of the time derivative on the left-hand side of the TDSE in Eq.~\eqref{eq:TDSE_int_pic_app}. The initial condition is not altered under this transformation, i.e., $\boldsymbol{c}_0'=\mathcal{T}\boldsymbol{c}_0=\boldsymbol{c}_0$.
Next, we diagonalize $\mathcal{H}'(t)$ to obtain the corresponding dressed state Hamiltonian. The eigenvalues of $\mathcal{H}'(t)$, describing the time-dependent eigenenergies of the dressed states, are given by
\begin{align}
	\lambda_0(t) & \equiv 0 \\
	\lambda_{\pm1}(t) & = \mp\hbar\sqrt{2\Omega_0^2(t)+\Delta^2(t)}=:\mp\hbar\Omega_g(t),
\end{align}
with the generalized Rabi frequency ${\Omega_g(t)=\sqrt{2\Omega_0^2(t)+\Delta^2(t)}}$. Defining the mixing angle $\Theta(t)$ through
\begin{equation}
	\tan[\Theta(t)]=\frac{\sqrt{2}\,\Omega_0(t)}{\Delta(t)},
\end{equation}
the unitary matrix
\begin{equation}\label{eq:trafo_dressed}
	\mathcal{D} = \begin{pmatrix}
	\cos(\Theta)	& \frac{1}{\sqrt{2}}\sin(\Theta) & \frac{1}{\sqrt{2}}\sin(\Theta) \\
	-\frac{1}{\sqrt{2}}\sin(\Theta)	& \cos^2\left(\frac{\Theta}{2}\right) & -\sin^2\left(\frac{\Theta}{2}\right) \\
	\frac{1}{\sqrt{2}}\sin(\Theta)	& \sin^2\left(\frac{\Theta}{2}\right) & -\cos^2\left(\frac{\Theta}{2}\right)
\end{pmatrix}
\end{equation}
diagonalizes $\mathcal{H}'(t)$ and yields the Hamiltonian in the dressed state picture as
\begin{align}
	\mathcal{H}_{ad} & =\mathcal{D}^\dagger\mathcal{H}'\mathcal{D}-i\hbar\mathcal{D}^\dagger\dot{\mathcal{D}} \\ & =-\hbar\begin{pmatrix}
		0	& \frac{i}{\sqrt{2}}\dot{\Theta} & \frac{i}{\sqrt{2}}\dot{\Theta} \\
		-\frac{i}{\sqrt{2}}\dot{\Theta}	& \Omega_g & 0 \\
		-\frac{i}{\sqrt{2}}\dot{\Theta}	& 0 & -\Omega_g
	\end{pmatrix}. \label{eq:hamiltonian_adiabatic}
\end{align} 
The initial condition is transformed to
\begin{equation}
	\boldsymbol{a}_0=\lim_{t\rightarrow-\infty}\left[\mathcal{D}^\dagger(t)\boldsymbol{c}_0'\right]=\begin{pmatrix}
		-1 \\
		0 \\
		0
	\end{pmatrix}.
\end{equation}
From Eq.~\eqref{eq:hamiltonian_adiabatic}, we derive the adiabatic condition as

\begin{equation}\label{eq:adia_cond}
	\frac{1}{\sqrt{2}}\,\big|\dot{\Theta}(t)\big|\ll\Omega_g(t),
\end{equation}
which reads explicitly
\begin{equation}\label{eq:adia_cond_expl}
	\big|\dot{\Omega}_0\Delta-\Omega_0\dot{\Delta}\big|\ll \left(2\Omega_0^2+\Delta^2\right)^{3/2}.
\end{equation}
Introducing the dimensionless parameter 
\begin{equation}\label{eq:alpha}
	\alpha(t)=\frac{\big|\dot{\Omega}_0\Delta-\Omega_0\dot{\Delta}\big|}{ \left(2\Omega_0^2+\Delta^2\right)^{3/2}},
\end{equation}
we obtain a quantitative measure for the transient adiabaticity of the interaction according to Eq.~\eqref{eq:adia_cond_expl}. If $\alpha(t)\ll 1$, then the interaction is adiabatic and the solution of the TDSE in the adiabatic picture remains constant in time with $\boldsymbol{a}(t)\equiv\boldsymbol{a}_0$. Inverse transformation into the instantaneous frequency picture then yields the bare state solution
\begin{equation}\label{eq:adia_sol}
	\boldsymbol{c}'(t)=\mathcal{D}(t)\boldsymbol{a}(t)=\frac{1}{\Omega_g(t)}\begin{pmatrix}
		-\Delta(t) \\
		\Omega_0(t) \\
		-\Omega_0(t)
	\end{pmatrix}.
\end{equation}
Finally, the solution in the interaction picture under adiabatic conditions is obtained as
\begin{equation}\label{eq:solution_inst_pic}
	\boldsymbol{c}(t)=\mathcal{T}^\dagger(t)\boldsymbol{c}'(t)=\frac{1}{\Omega_g(t)}\begin{pmatrix}
		-\Delta(t) \\
		\Omega^\ast(t) \\
		-\Omega(t)
	\end{pmatrix}.
\end{equation}
Expressed via the mixing angle, this result reads
\begin{equation}\label{eq:solution_int_pic}
	\boldsymbol{c}(t)=\frac{1}{\sqrt{2}}\begin{pmatrix}
		-\sqrt{2}\cos[\Theta(t)] \\
		\sin[\Theta(t)]\,e^{-i\zeta(t)} \\
		-\sin[\Theta(t)]\,e^{i\zeta(t)}
	\end{pmatrix}.
\end{equation}
The bare state solution described by Eqns.~\eqref{eq:solution_inst_pic} and \eqref{eq:solution_int_pic}, respectively, reveals explicitly $(i)$ that the ground state amplitude $c_0(t)$ is strictly real-valued throughout the interaction and $(ii)$ that the excited state amplitudes $c_{+1}(t)$ and $c_{-1}(t)$ are negative conjugates of one another, i.e., $c_{+1}(t)=-c_{-1}^\ast(t)$ for all times $t$. These properties are characteristic for the V-RAP scenario.
\begin{figure*}
	\includegraphics[width=\linewidth]{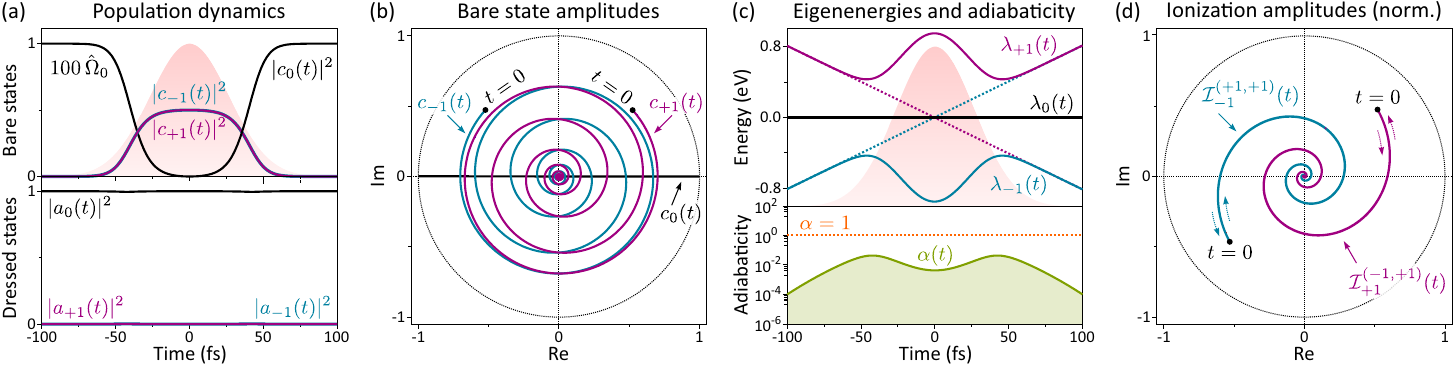}
	\caption{Simulated bound-state and ionization dynamics analogously to Fig.~\ref{fig2}, obtained for a peak Rabi frequency of $100\,\hat{\Omega}_0$, i.e., even further into the adiabatic regime.}\label{figA1}
\end{figure*}
\section{Ionization dynamics}\label{app:ionization_dynamics}
In this appendix, we utilize the adiabatic bare state solution from Eq.~\eqref{eq:solution_int_pic} to calculate the photoelectron amplitudes created by 2PI from the excited bound states using time-dependent perturbation theory. Perturbative 2PI from the states $|{+1}\rangle=|4p,+1\rangle$ and $|{-1}\rangle=|4p,-1\rangle$ creates photoelectron partial wave packets in the continua $|\varepsilon f,\pm1\rangle$ and $|\varepsilon f,\pm3\rangle$ along various 2PI pathways depicted in Fig.~\ref{fig1}(b). 
The 2PI pathways arise from the mixing of photons from the RCP field $\mathcal{E}_{-1}^-(t)$ and the LCP field $\mathcal{E}_{+1}^-(t)$, in different combinations. Each pathway is characterized by a vector $\boldsymbol{q}=(q_1,q_2)$, where the first (second) component describes the circularity ($\pm1$) of the first (second) absorbed photon.
Second order time-dependent perturbation theory predicts the photoelectron amplitude created from the bound excited state $|q_0\rangle$ along any 2PI pathway as
\begin{equation}\label{eq:photelectron_amp_general}
	a_{q_0}^{(q_1,q_2)}(\varepsilon)=\frac{\mu_{q_0}^{(q_1,q_2)}}{(i\hbar)^2}\int\limits_{-\infty}^\infty c_{q_0}(t)\mathcal{E}_{q_1}^-(t)\mathcal{E}_{q_2}^-(t)e^{i(\varepsilon-\varepsilon_0)\frac{t}{\hbar}}\,dt,
\end{equation}
with $\varepsilon_0=\varepsilon_{4p}+2\hbar\omega_0-\varepsilon_{IP}$ and $q_0,q_1,q_2=\pm1$. The prefactor $\mu_{q_0}^{(q_1,q_2)}$ is the effective two-photon dipole coupling for transition from the bound state $|q_0\rangle$ to the continuum, addressed via the pathway $\boldsymbol{q}=(q_1,q_2)$. Its angular part is determined by the product of the Clebsch-Gordan coefficients associated with the individual single-photon transitions, which are related to the Wigner $3j$-symbols:
\begin{align}
	\mu_{q_0}^{(q_1,q_2)} & \propto \left(\begin{array}{ccc}
	1	& 1 & 2 \notag\\
	q_0	& q_1 & -(q_0+q_1)
	\end{array}\right) \\
	& \quad\times\left(\begin{array}{ccc}
		2	& 1 & 3 \\
		q_0+q_1	& q_2 & -(q_0+q_1+q_2)
	\end{array}\right).
\end{align}
Assuming adiabatic interaction conditions in the bound system and making use of Eq.~\eqref{eq:solution_int_pic}, the excited state amplitudes are written explicitly as
\begin{equation}
	c_{q_0}(t) = \frac{q_0}{\sqrt{2}}\sin[\Theta(t)]e^{-iq_0\zeta(t)}.
\end{equation}
Together with the explicit form of the pulse components given in Eq.~\eqref{eq:env_chirp_comp}, the photoelectron amplitudes are then expressed as
\begin{equation}\label{eq:photelectron_amp_expl}
	a_{q_0}^{(q_1,q_2)}(\varepsilon)=\frac{1}{(i\hbar)^2}\int\limits_{-\infty}^\infty \mathcal{I}_{q_0}^{(q_1,q_2)}(t)e^{i(\varepsilon-\varepsilon_0)\frac{t}{\hbar}}\,dt,
\end{equation}
with the time-resolved and pathway-specific (1+2) REMPI amplitude
\begin{equation}
	\mathcal{I}_{q_0}^{(q_1,q_2)}(t)=\frac{q_0}{\sqrt{2}}\,\mu_{q_0}^{(q_1,q_2)}\sin[\Theta(t)]\mathcal{E}^2(t)e^{-i(q_0+q_1+q_2)\zeta(t)}.
\end{equation}
This result is the basis for our discussion of the adiabatic cancellation mechanism in Sec.~\eqref{sec:system:ionization} in the main text.
\bibliography{ultra_db.bib}
\end{document}